\documentclass[%
 aip,
 amsmath,amssymb,
 reprint,%
]{revtex4-1}

\usepackage{graphicx}
\usepackage{dcolumn}
\usepackage{bm}

\usepackage[utf8]{inputenc}
\usepackage[T1]{fontenc}
\usepackage{mathptmx}
\usepackage{etoolbox}
\usepackage[normalem]{ulem}
\usepackage{xcolor}
\usepackage{comment}
\usepackage{url}
\usepackage[
  colorlinks=true,
  urlcolor=blue,
  linkcolor=black,
  citecolor=black
]{hyperref}

\makeatletter
\def\@email#1#2{%
 \endgroup
 \patchcmd{\titleblock@produce}
  {\frontmatter@RRAPformat}
  {\frontmatter@RRAPformat{\produce@RRAP{*#1\href{mailto:#2}{#2}}}\frontmatter@RRAPformat}
  {}{}
}%
\makeatother
\begin{document}

\preprint{AIP/123-QED}

\title{Pt-induced semimetallic transport and enhanced thermoelectric response in the half-Heusler alloy YNi\textsubscript{0.5}Pt\textsubscript{0.5}Bi}


\author{Vikas Chahar}
\affiliation{ Department of Physics, Indian Institute of Technology Delhi, New Delhi 110016,~India
}%
 
\author{Shivam Rathod}%
\affiliation{ Department of Physics, Indian Institute of Technology Delhi, New Delhi 110016,~India
}%
\author{Lankun Wang}
\affiliation{State Key Laboratory of Precision Welding and Joining of Materials and Structures, Harbin Institute of Technology, Harbin, 150001, China
}
\author{Sneh}%
 \affiliation{ Department of Physics, Indian Institute of Technology Delhi, New Delhi 110016,~India
}%
\author{Shubhadeep Bej}%
\affiliation{ Department of Physics, Indian Institute of Technology Delhi, New Delhi 110016,~India
}%
\author{Divya Prakash Dubey}%
\affiliation{ Department of Microtechnology and Nanoscience, Chalmers University of Technology, SE-41296, Gothenburg, Sweden
}%


\author{Dinesh Dixit}%
\affiliation{ Central Research Facility, Indian Institute of Technology Delhi, New Delhi 110016,~India 
}%
\author{Abdelmoneim Elkhouly}%
\affiliation{Physics Department, Faculty of Science, Damanhour University, Damanhour 22516, Egypt
}%

\author{Raúl Cardoso-Gil}%
\affiliation{Max-Planck-Institut für Chemische Physik fester Stoﬀe, Nöthnitzer Str. 40, 01187 Dresden, Germany 
}%
\author{Zihang Liu}%
\affiliation{State Key Laboratory of Precision Welding and Joining of Materials and Structures, Harbin Institute of Technology, Harbin, 150001, China
}%

\author{Xiao Xu}%
\affiliation{Department of Materials Science, Graduate School of Engineering, Tohoku University, Sendai 980-8579, Japan
}%

\author{Rie Y. Umetsu}%
\affiliation{ Institute for Materials Research, Tohoku University, Sendai 980-8577,~Japan
}%
\altaffiliation[Also at ]{Center for Science and Innovation in Spintronics, Tohoku University, Sendai 980-8577, Japan}%

\author{Ratnamala Chatterjee}%
\affiliation{ Department of Physics, Indian Institute of Technology Delhi, New Delhi 110016,~India
}%

\author{Kaustuv Manna}
 \email{kaustuvmanna@iitd.ac.in}
 \homepage{https://web.iitd.ac.in/~kaustuvmanna/}
\affiliation{ Department of Physics, Indian Institute of Technology Delhi, New Delhi 110016,~India
}%


\begin{abstract}
\textcolor{black}{Half-Heusler alloys provide a versatile platform for tuning electronic and thermal transport through chemical substitution. Here, we investigate the thermoelectric and magnetotransport properties of spark-plasma-sintered YNi\textsubscript{0.5}Pt\textsubscript{0.5}Bi. Partial substitution of Ni by Pt drives YNiBi toward semimetallic transport, characterized by hole-dominated conduction, a low Hall carrier concentration of $\mathrm{1.9 \times 10^{19}  cm^{-3}}$, and a sizeable nonsaturating magnetoresistance of 18$\%$ at 300 K under 14 T. Compared with YNiBi, Pt alloying increases the room-temperature Seebeck coefficient from approximately 90 to 130 $\mathrm{\mu V K^{-1}}$, while reducing the thermal conductivity to 3.15 $\mathrm{W m^{-1} K^{-1}}$. Single-parabolic-band analysis indicates an enhanced carrier effective mass, resulting in a room-temperature power factor of 13.7 $\mathrm{\mu W cm^{-1} K^{-2}}$ and a figure of merit ($zT$= 0.13). First-principles calculations further show a Pt-induced closing of the electronic band gap and a low density of states near the Fermi level, consistent with the experimentally observed semimetal-like transport. These results demonstrate that Ni/Pt alloying provides an effective route for simultaneously tuning the electronic, magnetotransport, and thermoelectric properties of YNiBi-based half-Heusler compounds.}
\end{abstract}

\maketitle

\section{\label{sec:level1}Introduction} 
Thermoelectric materials and topologically non-trivial materials are of great interest in view of their potential applications. The clean energy needs, undesired generated heat in devices converted to useful energy, and an alternative source of energy is provided by the thermoelectric materials. Topologically non-trivial materials are one of the most promising materials for the future quantum technologies. With the application of advanced experimental techniques many materials have been successfully experimentally established as topologically non-trivial materials ~\cite{Liu2016,Hosen2020,doi:10.1126/science.1173034,Xia2009,Hasan2011}. Interestingly, many experimentally verified topologically non-trivial materials were also studied for their thermoelectric properties and found to be having good thermoelectric properties in comparison to all studied thermoelectric materials ~\cite{Xu2017,Park2021,Baldomir2019,Ivanov2018,Pal2015,Burkov2019,Pabst2019,Sattigeri2021,Venkatasubramanian2001,Huang2018,Mukhopadhyay2018}. In topologically non-trivial materials, a high z-element provides the spin-orbit coupling (SOC)~\cite{Laubach2014,Moore2010}. Good thermoelectric materials need lower thermal conductivity, where high z-element is useful to achieve the lower thermal conductivity~\cite{doi.org/10.1002/adma.200600527,Zhao2014}. The presence of dissipationless surface conducting paths in non-trivial materials can be useful for attaining good thermoelectric materials,~\cite{Ivanov2018} where good electrical conductivity and lower thermal conductivity are desired, but both are proportional to the charge carrier density and are challenging to control individually while increasing one and decreasing the other. Thus, many characteristics of a material responsible for its topologically non-trivial behavior are also required for good thermoelectric properties ~\cite{Xu2017,Baldomir2019,Ivanov2018,Park2021}.

Half-Heusler alloy is a class of materials having \textcolor{black}{$ABC$} chemical formula, where $A$ is the most electropositive element, $B$ is less electropositive transition element and $C$ is a p-block 
element, with cubic MgAgAs type non-centrosymmetric crystal structure and $F\bar{4}3m$ space group symmetry. In this class of materials, both topologically non-trivial alloys ~\cite{Huang2018,Singh2020,Ning2020,Nakajima2015,Wang2018,Xiao2010} and promising thermoelectric materials ~\cite{Quinn2021,Chauhan2023,Zhang2023,Ai2025,Mukhopadhyay2018,LI20241274,Zhu2023} have been found. Among various promising thermoelectric half-Heusler alloys, YNiBi is a potential p-type thermoelectric material~\cite{Sharma2018,Li2015,Shan2012}. It is a trivial half-Heusler alloy, with semiconducting and diamagnetic behavior. It has a moderate power factor of $\sim$ 11.4 $\mathrm{\mu W cm^{-1} K^{-2}}$ at 300 K, low lattice thermal conductivity, 3.5 $\mathrm{W m^{-1} K^{-1}}$,  which is favourable for a good thermoelectric material, Seebeck coefficient  90 $\mathrm{\mu V K^{-1}}$, and a figure of merit of $\sim$ 0.09  at 300 K ~\cite{Li2015}. It is well studied for its thermoelectric properties, but not reported for its magneto-resistance properties in literature. Among all topologically non-trivial half-Heusler alloys, YPtBi is a well established topologically non-trivial semimetal ~\cite{Kim2016,Kim2022,Liu2016,Shekhar2015,Bay2012,Butch2011,Hosen2020,10.1063/1.4799144}. It has hole dominated transport properties and is diamagnetic in nature. 
Additionally, it supports an unconventional spin 3/2 superconducting ground state~\cite{Kim2016}. ARPES studies of YPtBi confirm the presence of a Dirac cone at the $\Gamma$ point, located 500 meV below the Fermi energy~\cite{Hosen2020}. \textcolor{black}{Experimental thermoelectric characterization of bulk YPtBi remains comparatively limited}.

Considering that YNiBi is a promising thermoelectric half-Heusler alloy, whereas YPtBi is an established topologically non-trivial semimetal, alloying these two end members offers an attractive route to combine and optimize their desirable characteristics. The method of substitution, doping or alloying is well applied to materials to enhance or tune their thermoelectric properties~\cite{Mitra2022,Zhang2023,C4EE03042G} , and half-Heusler alloys are well adaptable to doping or alloying~\cite{Chadov2010,Casper2012} in literature to tune their properties.
~We here report the thermoelectric and magneto-transport properties of YNi\textsubscript{0.5}Pt\textsubscript{0.5}Bi alloy, where Pt and Ni are equally alloyed to occupy the $B$ site in the $ABC$ formula of the half-Heusler alloy, made by spark-plasma-sintering. The \textcolor{black}{average} nuclear charge, $\langle Z\rangle = \frac{1}{N}\sum_{i}^{} Z_i$, \textcolor{black}{where $N$ is 2 for binaries and 3 for ternaries, is used as an order parameter for half-Heusler alloys}, lies above 50 for a topological non-trivial half-Heusler alloy~\cite{Chadov2010}. \textcolor{black}{Average} nuclear charge is 50 for YNiBi, 66.6 for YPtBi and is 58.3 for YNi\textsubscript{0.5}Pt\textsubscript{0.5}Bi. The lattice parameter for YNiBi is 6.412 \AA, for YPtBi is 6.651 \AA. The tuning of lattice parameter by mixing Ni and Pt will optimise the hybridisation strength to bring the conditions for desired band inversion ~\cite{Chadov2010}. The band gap energy and the magnitude of spin orbit coupling are also supposed to be tuned by alloying Ni and Pt at $B$ site in $ABC$ formula of half-Heusler alloy to bring the desired conditions for possible topological non-triviality in the alloy.~\cite{Chadov2010} We have found that the thermoelectric properties in YNi\textsubscript{0.5}Pt\textsubscript{0.5}Bi are enhanced in comparison to the reported thermoelectric values for YNiBi~\cite{Li2015}.
We have also found 18$\%$  magnetoresistance ($MR$) at 14 T applied field at room temperature in YNi\textsubscript{0.5}Pt\textsubscript{0.5}Bi.    

\section{Experimental and First-principles calculation Details}
 Small chunks of high purity elements Y, Ni, and Pt are taken in the stoichiometric amounts and arc melted to form YNi\textsubscript{0.5}Pt\textsubscript{0.5} ingots. The ingots formed above are crushed in small pieces in a mortar pestle, weighed and then mixed with appropriate amounts of high purity small Bi pieces to get the stoichiometry of the sample as YNi\textsubscript{0.5}Pt\textsubscript{0.5}Bi. The mixture above formed is then ball milled in \textcolor{black}{high energy planetary ball mill} to get fine, thoroughly mixed powder. The process of arc melting is done inside the Ar gas glove box and ball milling is done in an inert Ar gas environment to avoid oxidation of the rare-earth element. The above obtained fine powder is used for spark-plasma-sintering at a pressure of 70 MPa and a temperature of 650 °C for obtaining 10 mm diameter and 2 mm thick pellets.  
The elemental composition of the pellet is determined using energy-dispersive X-ray spectroscopy (EDS) with a TESCAN Magna LMU system. The structural and phase characterization are carried out using powder X-ray diffraction (XRD) with Cu K$\alpha$ radiation with a Malvern PANalytical Empyrean diffractometer. 10 mm x 2 mm x 2 mm rectangular rods are cut from the pellet using a wire spark erosion cutter. These rectangular rods are polished and used for magneto-transport and thermoelectric measurements. The magneto-transport measurements are performed with a Cryogenic limited CFMS Physical Property Measurement System (PPMS) and a Quantum Design PPMS from 1.8 K to 300 K, at an applied magnetic field upto 14 T and 9 T, respectively. The thermoelectric measurements are performed with a Quantum Design PPMS from 10 K to 300 K. 

\textcolor{black}{
Density functional theory (DFT) calculations were performed using the Vienna ab initio simulation package (VASP)~\cite{PhysRevB.47.558,PhysRevB.54.11169,KRESSE199615,PhysRevB.49.14251} with the projector augmented wave method (PAW)~\cite{PhysRevB.50.17953,PhysRevB.59.1758}. The generalized gradient approximation of Perdew, Burke, and Ernzerhof (GGA-PBE)~\cite{PhysRevLett.77.3865} was used for the electronic exchange-correlation functional with an energy cutoff of 600 eV for the plane wave expansion. For lattice relaxation and structural optimization, the k-points were chosen as 8 × 8× 8 and the convergence accuracy of force was set as $10^{-3}$ eV/Å. The SOC effects were taken into account in the subsequent electronic structure calculations.}

\begin{figure*}[t]
  \centering

  \includegraphics[width=\textwidth]{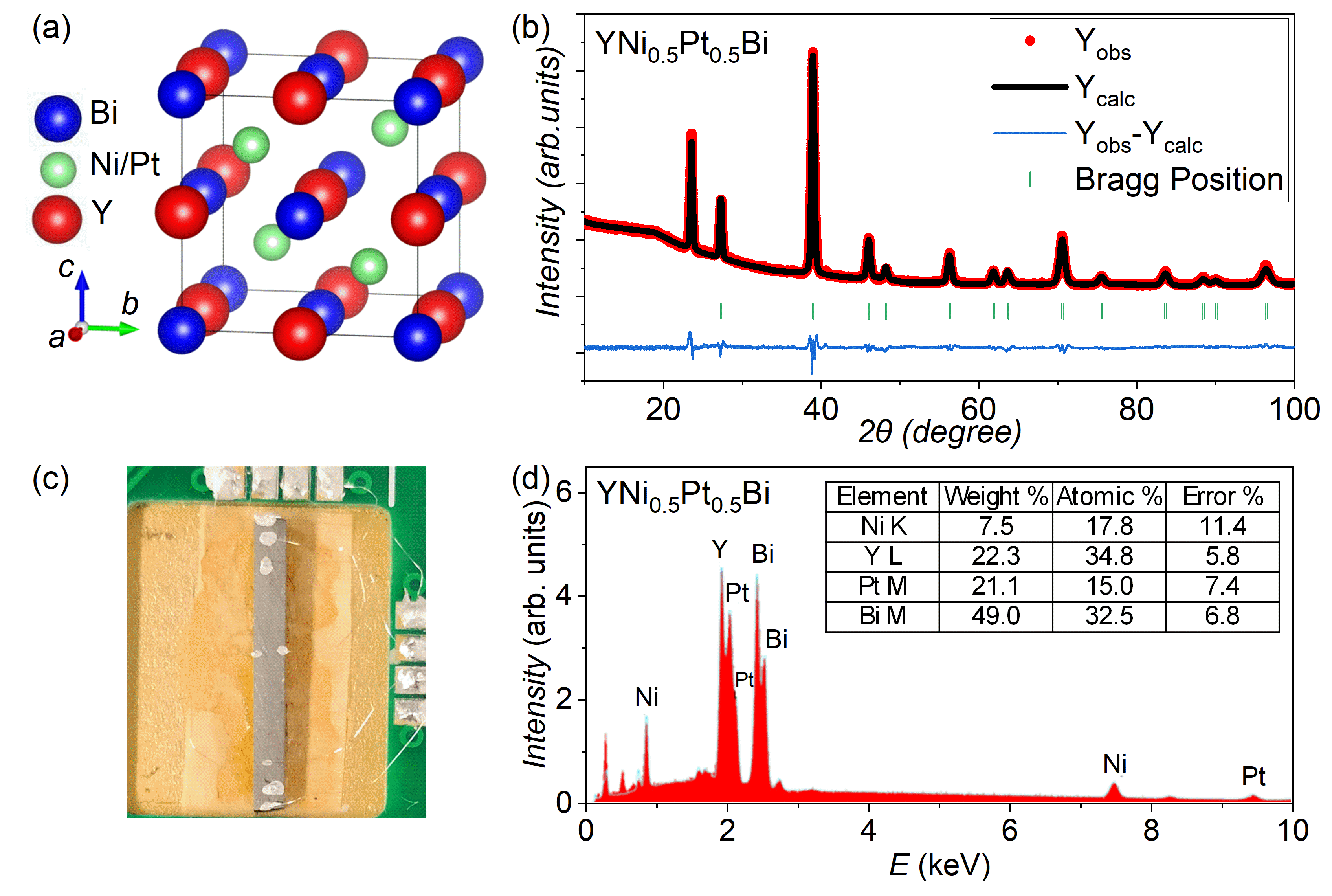}%

  \caption{\label{fig:one}
  (a) Crystal structure of Y(Ni/Pt)Bi. (b) Powder XRD pattern of pellet formed by spark-plasma-sintering (SPS) with its Rietveld refinement. (c) A picture of the rectangular sample rod obtained from the pellet via wire cutting, on which all electrical, magneto-transport and thermal transport measurements are performed. (d) Results of EDS measurement, performed on the finely ground pellet surface. The inset table shows the obtained atomic percentage of elements.}
\end{figure*}

\section{Results and Discussion}
\subsection{Sample characterization}

~Figure~\ref{fig:one}(a) illustrates the crystal structure of YNi\textsubscript{0.5}Pt\textsubscript{0.5}Bi, which crystallizes in the cubic MgAgAs type half-Heusler structure with space group \textit{F}$\bar{4}$3\textit{m}~(No. 216). The crystal structure consists of three interpenetrating face-centered cubic sublattices, each shifted relative to the others, along body diagonal by \textcolor{black}{one-fourth} of the body diagonal.

Figure~\ref{fig:one}~(b) shows the \textcolor{black}{Rietveld} refinement of powder XRD pattern, showing the compound is single phase. 
The lattice parameter of the cubic lattice obtained after the Rietveld refinement of the XRD pattern is  $a$ = 6.541 \AA. Figure~\ref{fig:one}~(c) depicts a picture of the bar shaped sample, mounted on PPMS measurement puck, used for the transport and thermoelectric measurements. Figure~\ref{fig:one}~(d) presents the EDS measurements results, performed on the surface of the pellet. The table in the inset of Fig.~\ref{fig:one}~(d) shows the atomic percentage (at. per.) of the four elements, which indicates nearly one-third composition of the Y (34.8 $\pm$ 2.02 at. per.) and Bi (32.5 $\pm$ 2.21 at. per.) elements and nearly one-third combined composition of Ni (17.8 $\pm$ 2.03 at. per.) and Pt (15.0 $\pm$ 1.11 at. per.), with approximately equal shares of Ni and Pt in composition.

\subsection{Electrical and magneto-transport measurements}

\begin{figure*}[t]
  \centering

  \includegraphics[width=\textwidth]{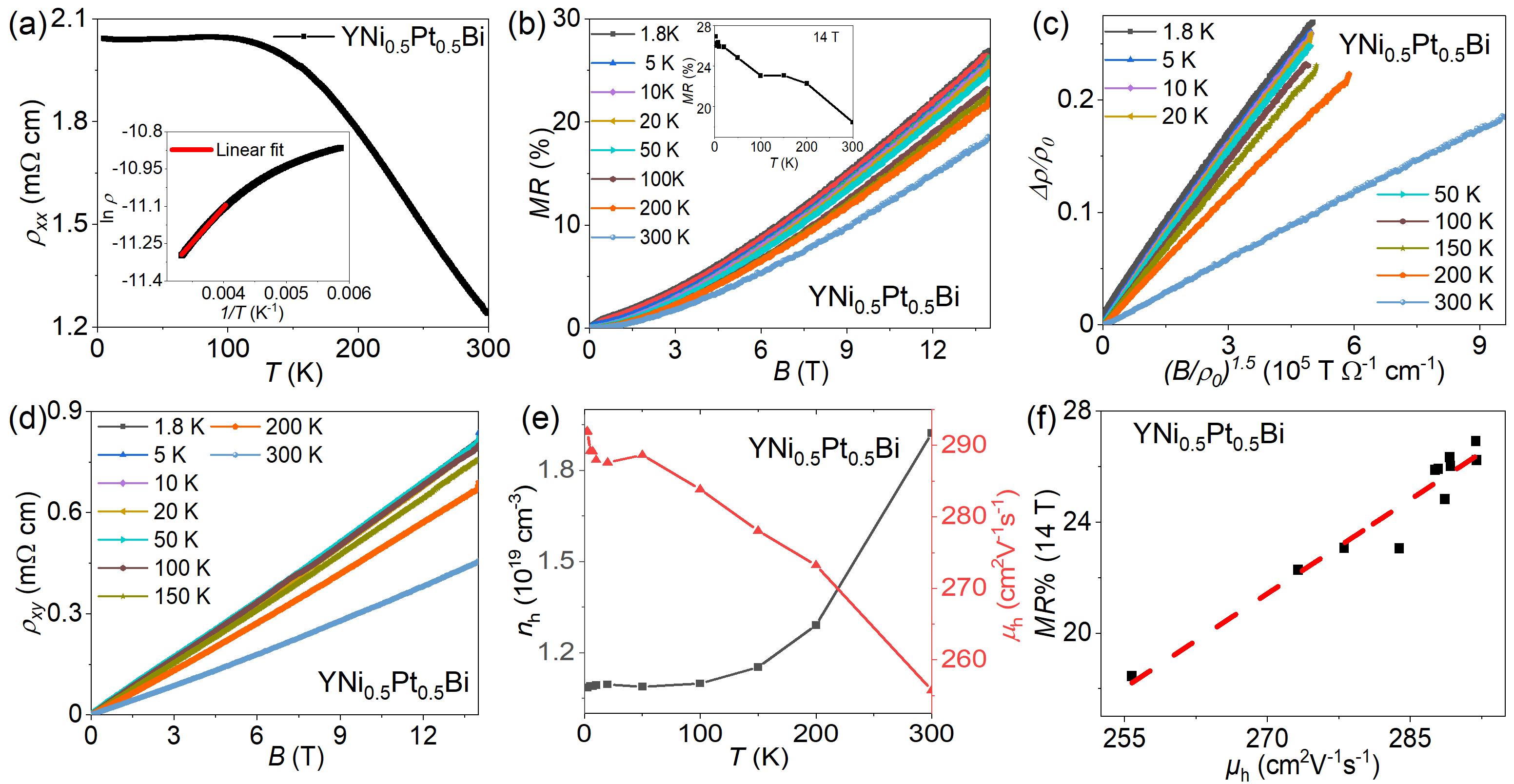}%

  \caption{\label{fig:two}
  (a) Resistivity dependence  with temperature from 5 K to 300 K. (b) Magnetoresistance percentage ($MR\%$) dependence with magnetic field applied perpendicular to the current, applied along the sample length, at different temperatures. (c) Kohler scaling of the $MR\%$ at different temperatures with $\Delta\rho/\rho_{0}$ versus $(B/\rho_{0})^{1.5}$. (d) \textcolor{black}{Hall resistivity dependence with applied magnetic field, at different temperatures. Current is applied along the sample rod length, magnetic field is applied perpendicular to the current direction, and Hall voltage is measured perpendicular to both current and applied magnetic field direction.} (e) Hole carrier concentration and hole mobility dependence with temperature. (f) $MR\%$ at 14 T at different temperatures plotted with the hole mobility at the corresponding temperature.}
\end{figure*}

~Figure~\ref{fig:two} (a) shows the 
temperature dependent resistivity
.  At low temperature, the resistivity increases slightly with increase in temperature upto $\sim$ 50 K, and then starts decreasing with increase in temperature and becomes nearly linear at higher temperatures above 250 K. 
The Arrhenius equation, \( \rho = \rho_0\,\exp\left(\frac{E_g}{2k_B T}\right) \) , can be used to calculate the energy band gap, $E_g$
. The $E_g$ calculated here from the Arrhenius fit, rather, represents the apparent activation energy extracted from the high-temperature resistivity, and is found to be 0.049 eV. The apparent activation energy is extracted from the slope of the plot between ln$\rho$ vs $1/T$ known as the Arrhenius plot, as shown in the inset of Fig.~\ref{fig:two} (a).
The temperature dependent behavior of resistivity for YNi\textsubscript{0.5}Pt\textsubscript{0.5}Bi is similar to temperature dependent behavior of resistivity for YPtBi ~\cite{Butch2011}.

~Figure~\ref{fig:two}~(b) shows the variation of $MR\%$ for YNi\textsubscript{0.5}Pt\textsubscript{0.5}Bi with applied magnetic fields at different temperatures. The magneto-resistance is measured upto 14 T, with magnetic field applied perpendicular to the current direction. At a high magnetic field of 14 T, $MR\%$ of 27$\%$ is obtained at 1.8 K, which is slightly decreasing with increase in temperature and becomes 18$\%$ at 300 K, as shown in the inset of Fig.~\ref{fig:two}~(b). In comparison to the perfectly linear behavior of $MR$ at a low temperature of 3 K upto an applied magnetic field of 14 T in YPtBi~\cite{Butch2011}, the $MR$ behavior in YNi\textsubscript{0.5}Pt\textsubscript{0.5}Bi deviates from linearity
. The Kohler scaling of the $MR\%$ at different temperatures showing a linear dependence of $\Delta\rho/\rho_{0}$ with $(B/\rho_{0})^{1.5}$ is shown in Fig.~\ref{fig:two}~(c). The $MR$ curves at different temperatures do not collapse onto a single curve, i.e. they do not follow Kohler's rule. It suggests the presence of multiple scattering mechanisms responsible for the magnetoresistance emergence. 

The Hall resistivity dependence with magnetic field at different temperatures is shown in Fig.~\ref{fig:two}~(d). The positive value of the Hall coefficient indicates the hole dominant transport in YNi\textsubscript{0.5}Pt\textsubscript{0.5}Bi. The hole charge carrier density and mobility calculated from the Hall data are shown in Fig.~\ref{fig:two}~(e). The low hole charge carrier density of 1.08$\times$$10^{19}$ $cm^{-3}$ at 2 K lies near the semimetallic regime~\cite{Shekhar2015}. The hole charge carrier density for YNi\textsubscript{0.5}Pt\textsubscript{0.5}Bi at 300 K is found to be 1.9 $\times$ $10^{19}$ $\mathrm{cm^{-3}}$ and hole mobility at 300 K is found to be 255.8 $\mathrm{cm^{2}V^{-1}s^{-1}}$. 
The hole charge carrier density, at 300 K, of YNi\textsubscript{0.5}Pt\textsubscript{0.5}Bi, YNiBi~\cite{Li2015} and YPtBi~\cite{Butch2011} are almost equal, 2.6 $\times$ $10^{19}$ $\mathrm{cm^{-3}}$ and 2 $\times$ $10^{19}$ $\mathrm{cm^{-3}}$ for YNiBi and YPtBi respectively. The hole charge carrier mobility, at 300 K, of YNi\textsubscript{0.5}Pt\textsubscript{0.5}Bi and YNiBi (254.8 $\mathrm{cm^{2}V^{-1}s^{-1}}$) ~\cite{Li2015} are also almost equal, while it is very less in \textcolor{black}{comparison} to the reported values of single crystalline YPtBi (1000 $\mathrm{cm^{2}V^{-1}s^{-1}}$)~\cite{Butch2011}. The temperature dependent resistivity behavior as presented in Fig.~\ref{fig:two}~(a), relates with the charge carrier density behavior with temperature presented in Fig.~\ref{fig:two}~(e). The hole charge carrier density is nearly constant upto $\sim$ 50 K. In the same temperature region, the resistivity value also has a small change with respect to the temperature. Figure~\ref{fig:two}~(f) shows the $MR\%$ at 14 T at different temperatures plotted with the hole mobility at the corresponding temperature. The decrease in $MR\%$ at 14 T with increase in temperature has a linear relationship with decrease in hole mobility with increase in temperature.

\begin{figure*}[t]
  \centering

  \includegraphics[width=\textwidth]{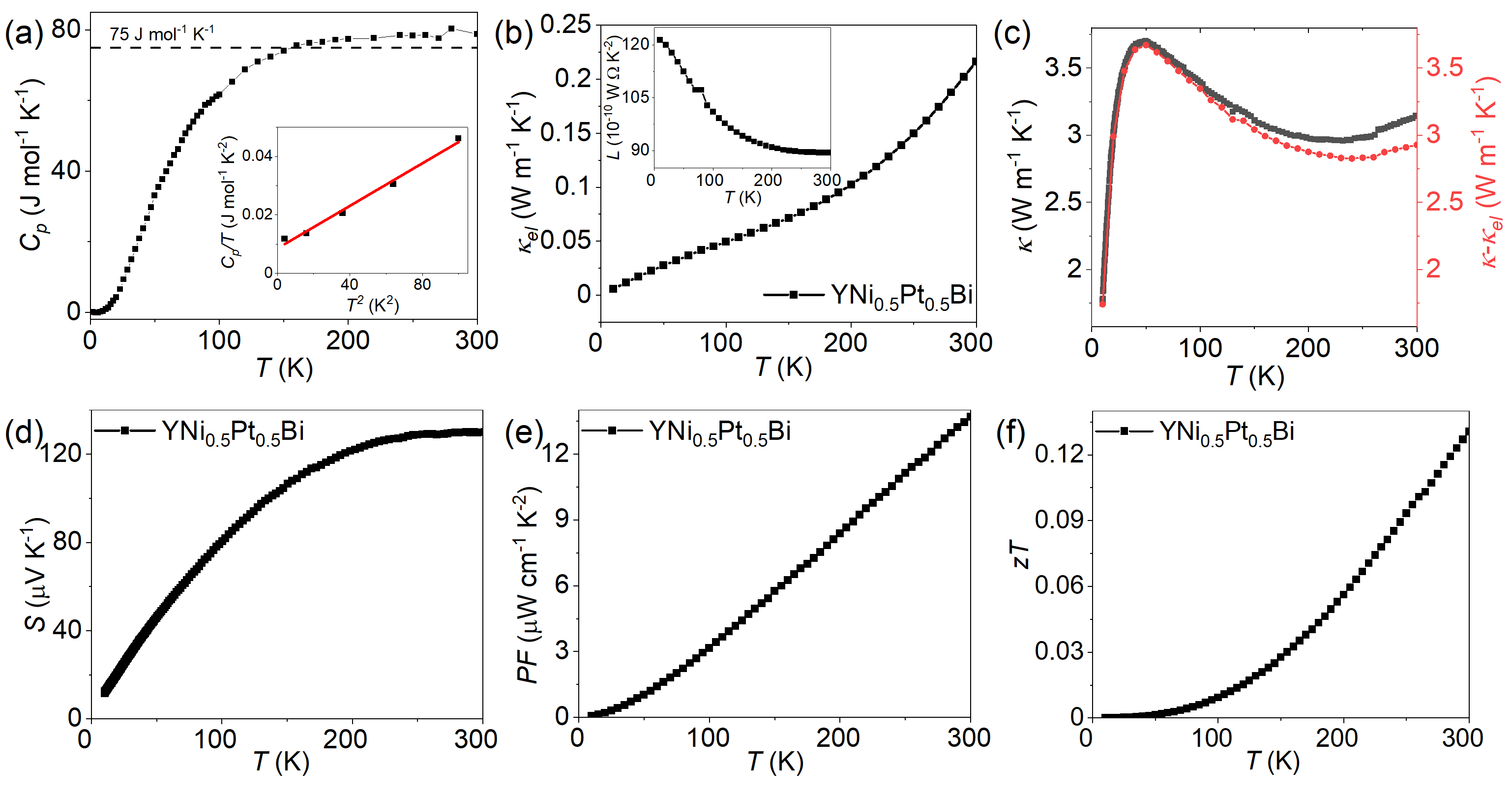}%

  \caption{\label{fig:three}
  (a) Heat capacity dependence with temperature. The inset figure shows the low temperature fitting of heat capacity using the equation $C_p/T = \gamma + \beta T^2$. (b) The temperature dependence of electronic contribution to the total thermal conductivity, computed via Wiedemann-Franz law. The graph in inset shows the Lorentz number, $L$, dependence with temperature, computed using the single-parabolic-band assumption. (c) Total thermal conductivity, $\kappa$, and $\kappa - \kappa_{el}$ dependence with temperature. (d) Seebeck coefficient, $S$, dependence with temperature. (e) Power factor, $PF$, dependence with temperature. (f) Figure of merit, $zT$, dependence with temperature.}
\end{figure*}

\begin{figure}[t]
  \centering

  \includegraphics[width=\columnwidth]{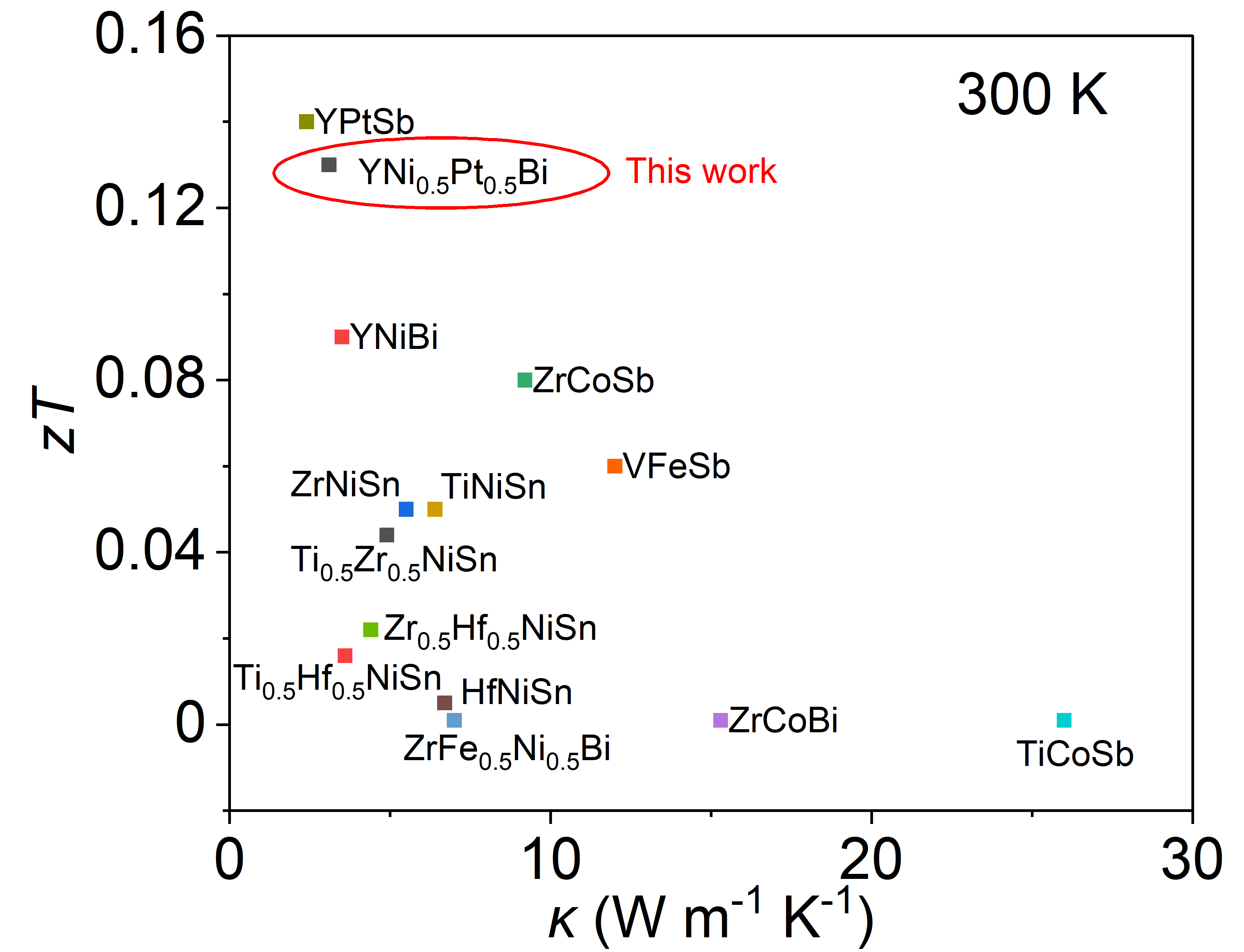}%

  \caption{\label{fig:four}
  $zT$ and thermal conductivity comparison plot of half-Heusler and double half-Heusler thermoelectric materials at 300 K as per the available data in literature.}
\end{figure}

\begin{table*}
\caption{\label{tab:table1} Lattice parameter, resistivity (at 5K and 300K), hole charge carrier density and hole mobility at 300 K, and $MR\%$ at 300 K and 9 T applied field for YNiBi, YNi\textsubscript{0.5}Pt\textsubscript{0.5}Bi and YPtBi.}
\begin{ruledtabular}
\begin{tabular}{lcccccccc}
Sample & $a$ (\AA) & \shortstack{$\rho_{\mathrm{xx(5K)}}$ \\$(\mathrm{m \Omega cm})$} & \shortstack{$\rho_{\mathrm{xx(300K)}}$ \\$(\mathrm{m \Omega cm})$} & \shortstack{$n_{\mathrm{h} \mathrm{(300K)}}$ \\$\mathrm{(cm^{-3})}$} & \shortstack{$\mu_{\mathrm{h} \mathrm{(300K)}}$\\ $\mathrm{(cm^{2}V^{-1}s^{-1})}$} & \shortstack{$MR\%$\\(9T, 300K)} & Sample form & Reference\\
\hline
YNiBi & $6.412$ & $\sim 1.9$ & $\sim$ 0.7 & 2.6 $\times 10^{19}$ & 254.8 & not reported & Polycrystalline & ~\citenum{Li2015}\\
YNi\textsubscript{0.5}Pt\textsubscript{0.5}Bi & 6.541 & 2.0 & 1.2 & 1.9 $\times 10^{19}$ & 255.8 & 9.85 & Polycrystalline & This work\\
YPtBi & 6.651 & 0.25 & 0.25 & 1.9 $\times 10^{19}$ & $\sim$1350 & 200 & Single crystal &~\citenum{Shekhar2015}\\
\end{tabular}
\end{ruledtabular}
\end{table*}

\begin{table*}
\caption{\label{tab:table2}Hole effective mass, thermal conductivity, Seebeck coefficient, power factor, figure of merit and energy band gap value of YNiBi and YNi\textsubscript{0.5}Pt\textsubscript{0.5}Bi at 300 K. 
}
\begin{ruledtabular}
\begin{tabular}{lcccccccc}
Sample & $ m^*_{\mathrm{300K}}$ & \shortstack{ $\kappa_\mathrm{300K}$  \\ $\mathrm{(Wm^{-1}K^{-1})}$ } & \shortstack{ $S_\mathrm{{300K}}$ \\ $\mathrm{({\mu}VK^{-1})}$ } & \shortstack{ $PF_\mathrm{{300K}}$ \\  $\mathrm{(\mu W cm^{-1}K^{-2})}$ } & $zT_\mathrm{{300K}}$ & \shortstack{$E_{g\mathrm{{300K}}} \mathrm{(eV)}$ \\ $\mathrm{(Arrhenius~ plot)}$ } &  Sample form &Reference\\
\hline
YNiBi & 0.40 $m_e$ & 3.5 & 90 & 11.4 & 0.09 & \textcolor{black}{0.31} & Polycrystalline & ~\citenum{Li2015}\\
YNi\textsubscript{0.5}Pt\textsubscript{0.5}Bi & 0.53 $m_e$ & 3.15 & 130 & 13.7 & 0.13 & 0.049 & Polycrystalline & This work\\
\end{tabular}
\end{ruledtabular}
\end{table*}

\subsection{Thermal transport measurements}
~Figure~\ref{fig:three}~(a) shows the specific heat dependence with temperature. With increase in temperature the specific heat increases and attains the Dulong-Petit limit of 75 $\mathrm{J mol^{-1} K^{-1}}$ near room temperature. The specific heat below 0.1 $\times$ $\theta_D$, where $\theta_D$ is the Debye temperature, can be described by the equation, $C_p = \gamma T + \beta T^3$. The term $\gamma T$ is due to electronic specific heat and $\beta T^3$ is due to phonon specific heat. The inset of Fig.~\ref{fig:three} (a) shows the linear fitting of $C_p/T$ versus $T^2$ plot, below 10 K, from where $\beta$ is found to be 0.36 \textcolor{black}{$\mathrm{mJ mol^{-1}K^{-4}}$}. $\beta$ is related with Debye temperature by the expression \textcolor{black}{$12\pi^4nR/5\theta_D^3$, where $n$ is the number of atoms per formula unit}, $R$ is the gas constant. Debye temperature is found to be \textcolor{black}{251.9} K for YNi\textsubscript{0.5}Pt\textsubscript{0.5}Bi. The reported value of $\theta_D$ for YNiBi is 219.9 K~\cite{Li2015}. The obtained value of Sommerfeld coefficient, $\gamma$ is $\lesssim$ 0.1 $\mathrm{Jmol^{-1}K^{-2}}$. The small value of $\gamma$ shows the lower contribution due to the electronic specific heat, which indicates a lower electronic DOS at \textcolor{black}{Fermi level}, consistent with small value of $\gamma$ for YPtBi~\cite{Pagliuso}.

Wiedemann-Franz law, $\kappa_{el} = LT\rho^{-1}$,~\cite{Jonson,Zhao2018} can be used to estimate the electronic part of thermal conductivity, where L is the Lorentz number. Lorentz number is calculated assuming the SPB model,~\cite{May2009} using the Eqs.\ref{eq:one}, \ref{eq:two} and \ref{eq:four} given below and is shown in the inset of Fig.~\ref{fig:three} (b). The electronic part of thermal conductivity, $\kappa_{el}$, estimated using above law is plotted in Fig.~\ref{fig:three} (b).

The electronic part of thermal conductivity, $\kappa_{el}$, is subtracted from the total thermal conductivity, $\kappa$, and is shown in Fig.~\ref{fig:three} (c), alongwith the total thermal conductivity, $\kappa$, with temperature. The $\kappa - \kappa_{el}$, is the sum of lattice thermal conductivity, $\kappa_{lat}$, and the bipolar contribution, $\kappa_b$, to the total thermal conductivity.
At low temperatures, the total thermal conductivity increases rapidly upto $\sim$ 45 K with increase in temperature. The sharp rise in thermal conductivity can be attributed to the sharp phonon population rise with increase in temperature at low temperatures. At 45 K, it reaches a local maxima and above 45 K, it decreases with increase in temperature, reaching a local minima at $\sim$ 230 K with a value of 2.96 Wm\textsuperscript{-1}K\textsuperscript{-1}. The decrease after sharp rise at low temperatures, is a typical behavior due to the Umklapp phonon scattering, due to which the thermal conductivity decreases with increase in temperature. With further increase in temperature, the thermal conductivity starts increasing, and at 300 K it attains a value of 3.15 Wm\textsuperscript{-1}K\textsuperscript{-1}. The rise in thermal conductivity while reaching 300 K can possibly be due to the involvement of diffusion processes due to presence of disorders in the system or due to rise in minority charge carrier (electronic) contribution to the thermal conductivity. The thermal conductivity of YNi\textsubscript{0.5}Pt\textsubscript{0.5}Bi at 300 K is slightly lower than the value of 3.5 Wm\textsuperscript{-1}K\textsuperscript{-1} reported for YNiBi at 300 K~\cite{Li2015}. 
The decrease in value of thermal conductivity of YNi\textsubscript{0.5}Pt\textsubscript{0.5}Bi at 300 K in comparison to YNiBi can possibly be due to rise in disorder in the lattice due to Ni and Pt mixing ~\cite{Liu2025} and due to addition of a relatively heavier element Pt ~\cite{Verma2023}. 

Figure~\ref{fig:three}~(d) shows the variation of Seebeck coefficient, $S$, with temperature. The positive value of Seebeck coefficient also suggest the dominance of p-type charge carrier, which is in agreement with the Hall analysis. The magnitude of $S$ increases with increase in temperature. $S$ increases from a value of $\sim$ 11.6 $\mu$VK\textsuperscript{-1} at 10 K to $\sim$ 130 $\mu$VK\textsuperscript{-1} at 300 K.

Assuming the single-parabolic-band assumption (SPB), the Seebeck coefficient, $S$, and Hall carrier density, $n_\mathrm{H}$, are used to calculate the carrier effective mass, $m^*$, at 300 K.\cite{May2009} The equations for Seebeck coefficient, Lorentz number, $L$, Hall carrier density, Fermi integral, $F_{i}(\eta)$ and Hall factor, $r_H$,~\cite{May2009} are given below, 

\begin{eqnarray}
S= \frac{{k}_{B}}{e}\left[\frac{(2+\lambda)F_{\lambda+1}(\eta)}{(1+\lambda)F_{\lambda}(\eta)}-\eta \right]\;
\label{eq:one},
\end{eqnarray}

\begin{eqnarray*}
L= \left( \frac{{k}_{B}}{e} \right)^{2}\frac{(1+\lambda)(3+\lambda)F_{\lambda}(\eta)F_{\lambda+2}(\eta)-(2+\lambda)^2F_{\lambda+1}(\eta)^2}{(1+\lambda)^2F_{\lambda}(\eta)^2} \;
\label{eq:two},
\end{eqnarray*}
\hfill(\ref{eq:two})

\addtocounter{equation}{1}

\begin{eqnarray}
n_H= {{4}}{{{\pi}}} \left(\frac{2m^*k_BT}{h^2} \right)^{\frac{3}{2}}\frac{F_{1/2}(\eta)}{r_H}
\label{eq:three},
\end{eqnarray}

\begin{eqnarray}
F_i(\eta) = \int_{0}^{\infty} \frac{\xi^i}{1+exp(\xi-\eta)} \,\mathrm{d}\xi\;,
\label{eq:four}
\end{eqnarray}

\begin{eqnarray}
r_H= \frac{3}{2}{F_{1/2}(\eta)} \left(\frac{({1/2}+2\lambda){F_{2\lambda-1/2}(\eta)}}{(1+\lambda)^2F_{\lambda}^2(\eta)} \right)
\label{eq:five},
\end{eqnarray}

\begin{figure*}[t]
  \centering

  \includegraphics[width=\textwidth]{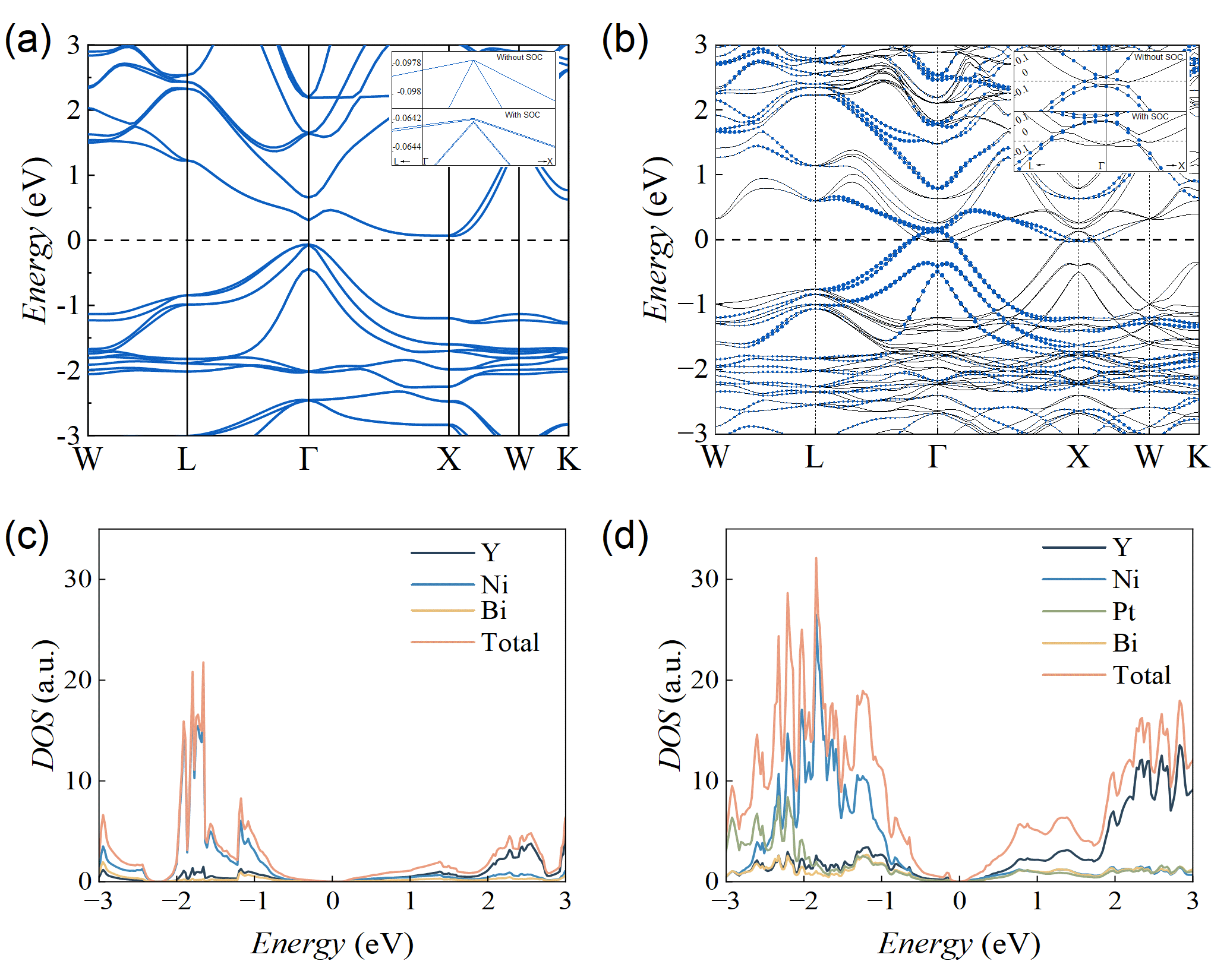}%

  \caption{\label{fig:five}
  (a) Electronic band structure of YNiBi \textcolor{black}{with SOC taken into account. The inset of (a) shows the VBM of YNiBi at $\Gamma$ in electronic band structure calculated with and without SOC taken into account}. (b) Electronic band structure of YNi\textsubscript{0.5}Pt\textsubscript{0.5}Bi \textcolor{black}{with SOC taken into account}. The size of blue dots depicts the contribution from the Pt orbitals. \textcolor{black}{The inset of (b) shows the VBM  and the CBM of YNi\textsubscript{0.5}Pt\textsubscript{0.5}Bi at $\Gamma$ in electronic band structure calculated with and without SOC taken into account}. (c) Partial and total density of states around Fermi level for YNiBi with SOC taken into account. (d) Partial and total density of states around Fermi level for YNi\textsubscript{0.5}Pt\textsubscript{0.5}Bi with SOC taken into account.}
\end{figure*}

where, $\xi$ is the reduced carrier energy, $\eta$ is the reduced Fermi energy, $k\textsubscript{B}$ is the Boltzmann constant, $h$ is the Planck constant, $e$ is the electron charge, and $\lambda$ is the scattering parameter.
The value of scattering parameter ($\lambda$ = 0) is used, assuming the carrier mobility to be limited by acoustic phonon scattering ~\cite{Goldsmid,May2009}. Assuming the carrier mobility limited by acoustic phonon scattering ($\lambda$ = 0), and using $S$ = 130 $\mu$VK\textsuperscript{-1} at 300 K in Eq. \ref{eq:one}, the obtained values are $\eta$ = 1.50, and $F_{1/2}(\eta)$ = 1.90. The Hall factor $r_H$ is calculated using Eq. \ref{eq:five} and obtained as 1.09. The effective mass of carrier, $m$\textsuperscript{*}, is obtained as 0.53 $m_e$ using Eq. \ref{eq:three} with $n_H$ = 1.9 $\times$ 10\textsuperscript{19} $cm\textsuperscript{-3}$ at 300 K.

 The Seebeck coefficient is much enhanced in comparison to the Seebeck coefficient reported as $\sim$ 90 $\mu$VK\textsuperscript{-1} at room temperature for YNiBi ~\cite{Li2015}. The Seebeck coefficient depends on the effective mass of charge carrier, $m^*$, and charge carrier concentration, $n$, by the relation, $S\propto m^*T/n^{2/3}$.~\cite{Levin2016,Jonson,Zhao2018} $m^*$ for YNi\textsubscript{0.5}Pt\textsubscript{0.5}Bi has increased and $n$ has decreased, which is contributing to the increase in Seebeck coefficient value with respect to YNiBi. The value of $S$ rises with increase in temperature and starts saturating near room temperature, which can be attributed to the increase in the minority charge carrier density at higher temperatures.  
The non-saturating magnetoresistance, low charge carrier density and low energy band gap value indicates the semimetal like nature of YNi\textsubscript{0.5}Pt\textsubscript{0.5}Bi. 

~Figure~\ref{fig:three}~(e) shows the power factor, $PF=S^2/\rho$, dependence with temperature. The \textit{PF} attains a maximum value of 13.7 $\mu Wcm^{-1}K^{-2}$ at 300 K, which is marginally higher to the value of 11.4 $\mu Wcm^{-1}K^{-2}$ reported for YNiBi ~\cite{Li2015} at 300 K. Figure~\ref{fig:three}~(f) shows the temperature dependence of figure of merit, $zT=S^2 T/\rho\kappa$, which attains value of 0.13 at 300 K. The value of $zT$ for YNi\textsubscript{0.5}Pt\textsubscript{0.5}Bi is enhanced in comparison to the reported value of 0.09 for YNiBi~\cite{Li2015} at 300 K. The increasing trend of dependence of $zT$ with temperature indicates that a higher figure of merit can still be obtained at higher temperatures above 300 K. Figure~\ref{fig:four} shows a comparison of $zT$ and thermal conductivities of half-Heusler and double Half-Heusler alloys according to the available data in literature ~\cite{Li2015,Min2024,Chauhan2019,Zhao2018,Barczak2018,Sekimoto2006,Heinrich_Hohl_1999,Ouardi2011,Chauhan2022,Kahiu2023} at 300 K.

\subsection{First-Principles Calculations}
\textcolor{black}{The experimental findings have suggested that for YNi\textsubscript{0.5}Pt\textsubscript{0.5}Bi the energy band gap has reduced with respect to YNiBi. In order to confirm this, we performed DFT-based first principles calculations. The crystal structure of YNi\textsubscript{0.5}Pt\textsubscript{0.5}Bi is $C$1$_b$ associated with $F\bar{4}3m$ space group. The unit cell contains 12 atoms, where atoms Bi, Y and Ni/Pt occupy (0,0,0), ($\frac{1}{2}$,$\frac{1}{2}$,$\frac{1}{2}$) and ($\frac{1}{4}$,$\frac{1}{4}$,$\frac{1}{4}$) \textcolor{black}{Wyckoff} positions~\cite{Mozharivskii1998}, respectively. In a unit cell composed of 12 atoms, the distances from a Ni-site atom to any of its nearest neighboring Ni-site atoms in the YNiBi system are all equal, and these nearest neighbors form a regular tetrahedron. Consequently, all possible Ni/Pt site occupancy configurations are symmetrically equivalent. Exchanging any Ni and Pt atom at their respective sites does not alter the structure.  
The electronic band structure and density of states (DOS) near Fermi level for YNiBi and YNi\textsubscript{0.5}Pt\textsubscript{0.5}Bi with SOC taken into account are shown in Fig.~\ref{fig:five}. The electronic band structure of YNiBi, plotted in Fig.~\ref{fig:five} (a), shows the presence of indirect energy band gap with valence band maxima (VBM) at $\Gamma$ and conduction band minima (CBM) at $\mathrm{X}$. 
}\textcolor{black}{The calculated band gap for YNiBi is found to be 0.13 eV. The earlier experimental and theoretical calculated values for YNiBi energy band gap are 0.31 eV~\cite{Li2015}, 0.248 eV~\cite{Li2015}, 0.13 eV~\cite{PhysRevB.59.15660}, 0.219 eV~\cite{https://doi.org/10.1002/adfm.200701369} and 0.26 eV~\cite{PhysRevLett.109.037602}.} 

\textcolor{black}{Figure~\ref{fig:five}~(b) demonstrates the electronic band structure of
YNi\textsubscript{0.5}Pt\textsubscript{0.5}Bi projected onto Pt orbitals. It \textcolor{black}{indicates} the closing of energy band gap in YNi\textsubscript{0.5}Pt\textsubscript{0.5}Bi. In YNi\textsubscript{0.5}Pt\textsubscript{0.5}Bi electronic band structure, the Fermi level crosses the valence band and conduction band at $\Gamma$ and $\mathrm{X}$ high symmetry locations. With respect to YNiBi, the VBM in YNi\textsubscript{0.5}Pt\textsubscript{0.5}Bi has shifted upward and the CBM has \textcolor{black}{shifted} downward, with higher Pt orbital projection in both VBM and CBM at $\Gamma$ and CBM at $\mathrm{X}$.} 
\textcolor{black}{In YNiBi electronic band structure calculated without SOC, valence band has three degenerate bands at VBM at $\Gamma$, two remain degenerate in $\Gamma-\mathrm{L}$ and $\Gamma-\mathrm{X}$ directions and one segregates in $\Gamma-\mathrm{L}$ and $\Gamma-\mathrm{X}$ directions, whereas in the electronic band structure calculated with SOC taken into account, all the three bands near VBM at $\Gamma$ are non-degenerate and each split in two bands, as shown in the inset of the Fig.~\ref{fig:five}~(a). The third band is not visible in zoomed inset of Fig.~\ref{fig:five}~(a) with SOC, but can be seen in Fig.~\ref{fig:five}~(a). Similarly, the one band at CBM at $\mathrm{X}$ in the electronic band structure calculated without SOC, splits in two when electronic band structure is calculated with SOC taken into account.
Also, in YNi\textsubscript{0.5}Pt\textsubscript{0.5}Bi, the CBM at $\Gamma$ has two energetically close lying bands when SOC is not taken into account, whereas with SOC taken into account, these two bands in CBM near Fermi level at $\Gamma$ split in two bands each, as shown in the inset of Fig.~\ref{fig:five}~(b). Also the three close valence bands near Fermi level in YNi\textsubscript{0.5}Pt\textsubscript{0.5}Bi at $\Gamma$ and $\mathrm{X}$ also split in two bands each, when SOC is taken into account in comparison to the electronic band structure calculated without SOC taken into account.} In YNi\textsubscript{0.5}Pt\textsubscript{0.5}Bi, the VBM at $\Gamma$ undergoes a curvature change from negative to positive and then again negative in the $\mathrm{L}-\Gamma-\mathrm{X}$ direction, with a small opening between the conduction band and valence band near the CBM and the VBM, just above the Fermi level, as shown in the lower inset of Fig.~\ref{fig:five} (b). 
The similar behavior is also shown by the CBM at $\Gamma$, where the curvature changes from positive to negative to positive in the same $\mathrm{L}-\Gamma-\mathrm{X}$ direction and also by the CBM and the VBM at $\mathrm{X}$. 

\textcolor{black}{Figure~\ref{fig:five} (c) shows the DOS for YNiBi, where both the partial DOS for each individual element and the total DOS have been depicted. It confirms the presence of an energy band gap near Fermi level. Figure~\ref{fig:five} (d) shows the partial and total DOS for YNi\textsubscript{0.5}Pt\textsubscript{0.5}Bi. It depicts the closing of energy band gap and a low charge carrier density at Fermi level, further validating the experimental findings of low Sommerfeld coefficient value, and supporting the semimetallic nature for YNi\textsubscript{0.5}Pt\textsubscript{0.5}Bi.}

\section{Conclusions}
%
In summary, polycrystalline YNi\textsubscript{0.5}Pt\textsubscript{0.5}Bi half-Heusler alloy was successfully synthesized by spark-plasma-sintering, and its thermoelectric and magnetotransport properties were systematically investigated. Pt alloying transforms the transport behavior towards a semimetallic state with hole-dominated conduction, a low hole carrier concentration of ($1.9\times10^{19}$ cm$^{-3}$), and a sizeable non-saturating magnetoresistance of (18$\%$) at 300 K under 14 T. \textcolor{black}{ The high-temperature apparent activation energy extracted using Arrhenius plot as a phenomenological simple activated model for YNi\textsubscript{0.5}Pt\textsubscript{0.5}Bi, is found to be 0.049 eV.} \textcolor{black}{The DFT based electronic band structure calculations shows the closing of energy band gap and a low charge carrier DOS at Fermi level. These calculations support the semimetallic nature of YNi\textsubscript{0.5}Pt\textsubscript{0.5}Bi.} The enhanced carrier effective mass together with the low carrier concentration results in an improved Seebeck coefficient of (130 $\mu$VK$^{-1}$), while alloy-disorder-induced phonon scattering reduces the thermal conductivity. Consequently, the room-temperature power factor and figure of merit reach 13.7 $\mu$Wcm$^{-1}$K$^{-2}$ and 0.13, respectively, representing a clear improvement over pristine YNiBi. Furthermore, the increasing trend of $zT$ at 300 K suggests that YNi\textsubscript{0.5}Pt\textsubscript{0.5}Bi has the potential to exhibit even better thermoelectric performance at higher temperatures. \textcolor{black}{The results of electronic band structure calculations of YNi\textsubscript{0.5}Pt\textsubscript{0.5}Bi} \textcolor{black}{ showing the modification of the VBM and the CBM at $\Gamma$ and $\mathrm{X}$}, ~\textcolor{black}{just above the Fermi level, suggests a possibility of band inversion.
}

\begin{acknowledgments}
K.M. acknowledges financial support from the Max Planck Society under the Max Planck--India Partner Group Project; the Science and Engineering Research Board (SERB), Department of Science and Technology (DST), Government of India, through Grant No.~CRG/2022/001826; the Aeronautics Research and Development Board (ARDB, Project No.~1992); and the Defence Research and Development Organisation (DRDO) under Project No.~DFTM/033203/P/41/JATC-P2QP17/10/D(R\&D)/2023. V.C. acknowledges the Ministry of Human Resource Development (MHRD) for Institute Fellowship support and funding through the GIMRT Program of the Institute for Materials Research, Tohoku University (Proposal No.~ 202308-CRKKE-0509). S.R. acknowledges the Anusandhan National Research Foundation (ANRF), India, for support through the National Postdoctoral Fellowship under Grant No.~PDF/2025/003558.
The authors also acknowledge the Central Research Facility (CRF) and the Nanoscale Research Facility at IIT Delhi for providing access to materials characterization facilities, including PPMS, MPMS, EDX, and XRD measurements.
\end{acknowledgments}

\nocite{*}
\bibliography{aipsamp}

@Article{C4EE03042G,
author ="Fu, Chenguang and Zhu, Tiejun and Liu, Yintu and Xie, Hanhui and Zhao, Xinbing",
title  ="Band engineering of high performance p-type {FeNbSb} based half-{{Heusler}} thermoelectric materials for figure of merit \textit{z{T}} > 1",
journal  ="Energy Environ. Sci.",
year  ="2015",
volume  ="8",
issue  ="1",
pages  ="216-220",
publisher  ="The Royal Society of Chemistry",
doi  ="10.1039/C4EE03042G",
url  ="http://dx.doi.org/10.1039/C4EE03042G"}

@article{Li2015,
   author = {Shanming Li and Huaizhou Zhao and Dandan Li and Shifeng Jin and Lin Gu},
   doi = {10.1063/1.4921811},
   issn = {10897550},
   issue = {20},
   journal = {J. Appl. Phys.},
   month = {5},
   publisher = {American Institute of Physics Inc.},
   title = {Synthesis and thermoelectric properties of half-{{Heusler}} alloy {YNiBi}},
   volume = {117},
   pages = {205101},
   year = {2015}
}

@article{Baldomir2019,
   author = {Daniel Baldomir and Daniel Faílde},
   doi = {10.1038/s41598-019-42744-3},
   issn = {20452322},
   issue = {1},
   journal = {Sci. Rep.},
   title = {On Behind the Physics of the Thermoelectricity of Topological Insulators},
   volume = {9},
   pages = {6324},
   year = {2019}
}

@article{Ivanov2018,
   author = {Yuri V. Ivanov and Alexander T. Burkov and Dmitry A. Pshenay-Severin},
   doi = {10.1002/pssb.201800020},
   issn = {15213951},
   issue = {7},
   journal = {Phys. Status Solidi B},
   title = {Thermoelectric Properties of Topological Insulators},
   volume = {255},
   pages = {1800020},
   year = {2018}
}

@article{Park2021,
   author = {Junsoo Park and Yi Xia and Vidvuds Ozoliņš and Anubhav Jain},
   doi = {10.1038/s41524-021-00512-w},
   issn = {20573960},
   issue = {1},
   journal = {npj Comput. Mater.},
   title = {Optimal band structure for thermoelectrics with realistic scattering and bands},
   volume = {7},
   pages = {43},
   year = {2021}
}

@article{Xu2017,
   author = {Ning Xu and Yong Xu and Jia Zhu},
   doi = {10.1038/s41535-017-0054-3},
   issn = {23974648},
   issue = {1},
   journal = {npj Quantum Mater.},
   title = {Topological insulators for thermoelectrics},
   volume = {2},
   pages = {51},
   year = {2017}
}

@article{Pal2015,
   author = {Koushik Pal and Shashwat Anand and Umesh V. Waghmare},
   doi = {10.1039/c5tc02344k},
   issn = {20507526},
   issue = {46},
   journal = {J. Mater. Chem. C},
   title = {Thermoelectric properties of materials with nontrivial electronic topology},
   volume = {3},
   number = {46},
   pages = {12130–12139},
   year = {2015}
}

@article{Burkov2019,
   author = {A. T. Burkov and Yu V. Ivanov and K. Nielsch and S. V. Novikov and N. R. Perez and D. A. Pshenay-Severin and H. Reith and L. U. Schnatmann and G. Schierning and M. P. Volkov},
   doi = {10.1016/j.matpr.2019.02.051},
   issn = {22147853},
   booktitle = {Mater. Today: Proc.},
   title = {Thermoelectric properties of silicides with topologically non-trivial electronic structure: {Co\textsubscript{1-x}M\textsubscript{x}Si (M=Fe, Ni)}},
   volume = {8},
   pages = {540-545},
   year = {2019}
}

@article{Pabst2019,
   author = {Falk Pabst and Dean Hobbis and Noha Alzahrani and Hsin Wang and I. P. Rusinov and E. V. Chulkov and Joshua Martin and Michael Ruck and George S. Nolas},
   doi = {10.1063/1.5116369},
   issn = {10897550},
   issue = {10},
   journal = {J. Appl. Phys.},
   title = {Transport properties of topologically non-trivial bismuth tellurobromides {Bi\textsubscript{n}TeBr}},
   volume = {126},
   pages = {105105},
   year = {2019}
}

@article{Sattigeri2021,
   author = {Raghottam M. Sattigeri and Trupti K. Gajaria and Prafulla K. Jha and Piotr Śpiewak and Krzysztof J. Kurzydłowski},
   doi = {10.1088/1361-648X/abdce8},
   issn = {1361648X},
   issue = {15},
   journal = {J. Phys.: Condens. Matter},
   title = {Emergence of -s, -p-d band inversion in zincblende gold iodide topological insulator and its thermoelectric properties},
   volume = {33},
   pages = {155402},
   year = {2021}
}

@article{Venkatasubramanian2001,
   author = {R. Venkatasubramanian and E. Siivola and T. Colpitts and B. O'Quinn},
   doi = {10.1038/35098012},
   issn = {00280836},
   issue = {6856},
   journal = {Nature},
   title = {Thin-film thermoelectric devices with high room-temperature figures of merit},
   volume = {413},
   pages = {597–602},
   year = {2001}
}

@article{Huang2018,
   author = {Shan Huang and Xiaowei Liu and Wenwen Zheng and Jingjing Guo and Rui Xiong and Ziyu Wang and Jing Shi},
   doi = {10.1039/c8ta07350c},
   issn = {20507496},
   issue = {41},
   journal = {J. Mater. Chem. A},
   title = {Dramatically improving thermoelectric performance of topological half-{{Heusler}} compound {LuPtSb}: Via hydrostatic pressure},
   volume = {6},
   pages = {20069–20075},
   year = {2018}
}

@article{Mukhopadhyay2018,
   author = {A. Mukhopadhyay and N. Lakshminarasimhan and N. Mohapatra},
   doi = {10.1088/1361-6463/aac567},
   issn = {13616463},
   issue = {26},
   journal = {J. Phys. D: Appl. Phys.},
   title = {Multi-functional properties of non-centrosymmetric ternary half-{{Heusler}s}, {RPdSb (R = Er and Ho)}},
   volume = {51},
   pages = {265004},
   year = {2018}
}

@article{Ai2025,
   author = {Xin Ai and Yu Wu and Haiyan Lyu and Lars Giebeler and Wenhua Xue and Andrei Sotnikov and Yumei Wang and Qihao Zhang and Denys Makarov and Yuan Yu and G. Jeffrey Snyder and Kornelius Nielsch and Ran He},
   doi = {10.1038/s41467-025-61868-x},
   issn = {20411723},
   issue = {1},
   journal = {Nat. Commun.},
   month = {12},
   pmid = {40659688},
   publisher = {Nature Research},
   title = {High-performance {ZrNiSn}-based half-{{Heusler}} thermoelectrics with hierarchical architectures enabled by reactive sintering},
   volume = {16},
   pages = {6497},
   year = {2025}
}

@article{Quinn2021,
   author = {Robert J. Quinn and Jan Willem G. Bos},
   doi = {10.1039/d1ma00707f},
   issn = {26335409},
   issue = {19},
   journal = {Mater. Adv.},
   month = {10},
   pages = {6246-6266},
   publisher = {Royal Society of Chemistry},
   title = {Advances in half-{{Heusler}} alloys for thermoelectric power generation},
   volume = {2},
   year = {2021}
}

@article{Chauhan2023,
   author = {Nagendra S. Chauhan and Yuzuru Miyazaki},
   doi = {10.1002/cnma.202200403},
   issn = {2199692X},
   issue = {1},
   journal = {ChemNanoMat},
   month = {1},
   publisher = {John Wiley and Sons Inc},
   title = {Iron-based Semiconducting Half-{{Heusler}} Alloys for Thermoelectric Applications},
   volume = {9},
   pages = {e202200403},
   year = {2023}
}

@article{Zhang2023,
   author = {Xiaoling Zhang and Ming Huang and Hongjun Li and Jiaxin Chen and Pengfei Xu and Biao Xu and Yifeng Wang and Guodong Tang and Sen Yang},
   doi = {10.1039/d3ta00631j},
   issn = {20507496},
   issue = {15},
   journal = {J. Mater. Chem. A},
   month = {3},
   pages = {8150-8161},
   publisher = {Royal Society of Chemistry},
   title = {Ultralow lattice thermal conductivity and improved thermoelectric performance in a {Hf}-free half-{{Heusler}} compound modulated by entropy engineering},
   volume = {11},
   year = {2023}
}

@article{Singh2020,
   author = {Ashutosh Kumar Singh and S. D. Ramarao and Sebastian C. Peter},
   doi = {10.1063/5.0006118},
   issn = {2166532X},
   issue = {6},
   journal = {APL Mater.},
   month = {6},
   publisher = {American Institute of Physics Inc.},
   title = {Rare-earth based half-{{Heusler}} topological quantum materials: {A} perspective},
   volume = {8},
   pages = {060903},
   year = {2020}
}

@article{Ning2020,
   author = {Suiting Ning and Shan Huang and Ziye Zhang and Renqi Zhang and Ning Qi and Zhiquan Chen},
   doi = {10.1039/d0cp01442g},
   issn = {14639076},
   issue = {26},
   journal = {Phys. Chem. Chem. Phys.},
   month = {7},
   pages = {14621-14629},
   pmid = {32567608},
   publisher = {Royal Society of Chemistry},
   title = {High thermoelectric performance of topological half-{{Heusler}} compound LaPtBi achieved by hydrostatic pressure},
   volume = {22},
   year = {2020}
}

@article{Nakajima2015,
   author = {Yasuyuki Nakajima and Rongwei Hu and Kevin Kirshenbaum and Alex Hughes and Paul Syers and Xiangfeng Wang and Kefeng Wang and Renxiong Wang and Shanta R. Saha and Daniel Pratt and Jeffrey W. Lynn and Johnpierre Paglione},
   doi = {10.1126/sciadv.1500242},
   issn = {23752548},
   issue = {5},
   journal = {Sci. Adv.},
   month = {6},
   publisher = {American Association for the Advancement of Science},
   title = {Topological RPdBi half-{{Heusler}} semimetals: A new family of noncentrosymmetric magnetic superconductors},
   volume = {1},
   pages = {e1500242},
   year = {2015}
}

@article{Wang2018,
   author = {Qing Ze Wang and Jiabin Yu and Chao Xing Liu},
   doi = {10.1103/PhysRevB.97.224507},
   issn = {24699969},
   issue = {22},
   journal = {Phys. Rev. B},
   month = {6},
   publisher = {American Physical Society},
   title = {Unconventional superconductivity and surface pairing symmetry in half-{{Heusler}} compounds},
   volume = {97},
   pages = {224507},
   year = {2018}
}

@article{Xiao2010,
   author = {Di Xiao and Yugui Yao and Wanxiang Feng and Jun Wen and Wenguang Zhu and Xing Qiu Chen and G. Malcolm Stocks and Zhenyu Zhang},
   doi = {10.1103/PhysRevLett.105.096404},
   issn = {00319007},
   issue = {9},
   journal = {Phys. Rev. Lett.},
   month = {8},
   title = {Half-{{Heusler}} compounds as a new class of three-dimensional topological insulators},
   volume = {105},
   pages = {096404},
   year = {2010}
}

@article{Kim2016,
   author = {Hyunsoo Kim and Kefeng Wang and Yasuyuki Nakajima and Rongwei Hu and Steven Ziemak and Paul Syers and Limin Wang and Halyna Hodovanets and Jonathan D. Denlinger and Philip M. R. Brydon and Daniel F. Agterberg and Makariy A. Tanatar and Ruslan Prozorov and Johnpierre Paglione},
   doi = {10.1126/sciadv.aao4513},
   month = {3},
   title = {Beyond Triplet: Unconventional Superconductivity in a Spin-3/2 Topological Semimetal},
   url = {http://arxiv.org/abs/1603.03375 http://dx.doi.org/10.1126/sciadv.aao4513},
   journal = {Sci. Adv.},
   volume = {4},
   pages = {eaao4513},
   year = {2016}
}

@article{Hosen2020,
   author = {M. Mofazzel Hosen and Gyanendra Dhakal and Klauss Dimitri and Hongchul Choi and Firoza Kabir and Christopher Sims and Orest Pavlosiuk and Piotr Wiśniewski and Tomasz Durakiewicz and Jian Xin Zhu and Dariusz Kaczorowski and Madhab Neupane},
   doi = {10.1038/s41598-020-69284-5},
   issn = {20452322},
   issue = {1},
   journal = {Sci. Rep.},
   title = {Observation of Dirac state in half-{{Heusler}} material {YPtBi}},
   volume = {10},
   pages = {12343},
   year = {2020}
}

@article{Shekhar2015,
   author = {Chandra Shekhar and Erik Kampert and Tobias Foerster and Binghai Yan and Ajaya K. Nayak and Michael Nicklas and Claudia Felser},
   month = {2},
   title = {Large linear magnetoresistance and weak anti-localization in {Y(Lu)PtBi} topological insulators},
   url = {http://arxiv.org/abs/1502.00604},
   journal = {arXiv:1502.00604},
   year = {2015}
}

@article{Sharma2018,
   author = {Sonu Sharma and Pradeep Kumar},
   doi = {10.1088/2053-1591/aabbf1},
   issn = {20531591},
   issue = {4},
   journal = {Mater. Res. Express},
   month = {4},
   publisher = {Institute of Physics Publishing},
   title = {Tuning the thermoelectric properties of {YNiBi} half-{{Heusler}} alloy},
   volume = {5},
   pages = {046528},
   year = {2018}
}

@article{Shan2012,
   author = {Rong Shan and Siham Ouardi and Gerhard H. Fecher and Li Gao and Andrew Kellock and Andrei Gloskovskii and Carlos E. Violbarbosa and Eiji Ikenaga and Claudia Felser and Stuart S.P. Parkin},
   doi = {10.1063/1.4764520},
   issn = {00036951},
   issue = {21},
   journal = {Appl. Phys. Lett.},
   month = {11},
   title = {A p-type {Heusler} compound: Growth, structure, and properties of epitaxial thin {NiYBi} films on {MgO}(100)},
   volume = {101},
   pages = {212102},
   year = {2012}
}

@article{Kim2022,
   author = {Hyunsoo Kim and Junhyun Lee and Halyna Hodovanets and Kefeng Wang and Jay D. Sau and Johnpierre Paglione},
   doi = {10.1103/PhysRevResearch.4.033169},
   issn = {26431564},
   issue = {3},
   journal = {Phys. Rev. Research},
   month = {7},
   publisher = {American Physical Society},
   title = {Quantum oscillations of the j=3/2 Fermi surface in the topological semimetal {YPtBi}},
   volume = {4},
   pages = {033169},
   year = {2022}
}

@article{Liu2016,
   author = {Z. K. Liu and L. X. Yang and S. C. Wu and C. Shekhar and J. Jiang and H. F. Yang and Y. Zhang and S. K. Mo and Z. Hussain and B. Yan and C. Felser and Y. L. Chen},
   doi = {10.1038/ncomms12924},
   issn = {20411723},
   journal = {Nat. Commun.},
   month = {9},
   publisher = {Nature Publishing Group},
   title = {Observation of unusual topological surface states in half-{Heusler} compounds {LnPtBi (Ln=Lu, Y)}},
   volume = {7},
   pages = {12924},
   year = {2016}
}

@article{Bay2012,
   author = {T. V. Bay and T. Naka and Y. K. Huang and A. De Visser},
   doi = {10.1103/PhysRevB.86.064515},
   issn = {10980121},
   issue = {6},
   journal = {Phys. Rev. B},
   month = {8},
   title = {Superconductivity in noncentrosymmetric {YPtBi} under pressure},
   volume = {86},
   pages = {064515},
   year = {2012}
}

@article{Butch2011,
   author = {N. P. Butch and P. Syers and K. Kirshenbaum and A. P. Hope and J. Paglione},
   doi = {10.1103/PhysRevB.84.220504},
   issn = {10980121},
   issue = {22},
   journal = {Phys. Rev. B},
   month = {12},
   title = {Superconductivity in the topological semimetal {YPtBi}},
   volume = {84},
   pages = {220504(R)},
   year = {2011}
}

@article{May2009,
   author = {Andrew F. May and Eric S. Toberer and Ali Saramat and G. Jeffrey Snyder},
   doi = {10.1103/PhysRevB.80.125205},
   issn = {10980121},
   issue = {12},
   journal = {Phys. Rev. B},
   month = {9},
   title = {Characterization and analysis of thermoelectric transport in n -type {Ba\textsubscript{8}Ga\textsubscript{16-x}Ge\textsubscript{30+x}}},
   volume = {80},
   pages = {125205},
   year = {2009}
}

@BOOK{Goldsmid,
   author       = {H. J. Goldsmid},
   year         = 2010,
   title        = {Introduction to Thermoelectricity},
   series       = {Springer Series in Materials Science},
   publisher    = {Springer}
}

@article{LI20241274,
title = {Half-{Heusler} thermoelectrics: Advances from materials fundamental to device engineering},
journal = {Joule},
volume = {8},
number = {5},
pages = {1274-1311},
year = {2024},
issn = {2542-4351},
doi = {https://doi.org/10.1016/j.joule.2024.03.016},
url = {https://www.sciencedirect.com/science/article/pii/S254243512400151X},
author = {Wenjie Li and Subrata Ghosh and Na Liu and Bed Poudel}
}

@article{Jonson,
   author = {M Jonson~ and D Mahan},
   journal = {Phys. Rev. B},
   title = {Mott's formula for the thermopower and the {Wiedemann-Franz law}},
   doi = {10.1103/PhysRevB.21.4223},
   volume = {21},
   pages = {4223},
   year = {1980}
}

@article{
doi:10.1126/science.1173034,
author = {Y. L. Chen  and J. G. Analytis  and J.-H. Chu  and Z. K. Liu  and S.-K. Mo  and X. L. Qi  and H. J. Zhang  and D. H. Lu  and X. Dai  and Z. Fang  and S. C. Zhang  and I. R. Fisher  and Z. Hussain  and Z.-X. Shen },
title = {Experimental Realization of a Three-Dimensional Topological Insulator, {Bi\textsubscript{2}Te\textsubscript{3}}},
journal = {Science},
volume = {325},
number = {5937},
pages = {178-181},
year = {2009},
doi = {10.1126/science.1173034},
URL = {https://www.science.org/doi/abs/10.1126/science.1173034}
}

@article{Xia2009,
   author = {Y. Xia and D. Qian and D. Hsieh and L. Wray and A. Pal and H. Lin and A. Bansil and D. Grauer and Y. S. Hor and R. J. Cava and M. Z. Hasan},
   doi = {10.1038/nphys1274},
   issn = {17452481},
   issue = {6},
   journal = {Nat. Phys.},
   pages = {398-402},
   publisher = {Nature Publishing Group},
   title = {Observation of a large-gap topological-insulator class with a single Dirac cone on the surface},
   volume = {5},
   year = {2009}
}

@article{Hasan2011,
   author = {M. Zahid Hasan and Joel E. Moore},
   doi = {10.1146/annurev-conmatphys-062910-140432},
   issn = {1947-5454},
   issue = {1},
   journal = {Annu. Rev. Condens. Matter Phys.},
   month = {3},
   pages = {55-78},
   publisher = {Annual Reviews},
   title = {Three-Dimensional Topological Insulators},
   volume = {2},
   year = {2011}
}

@article{Laubach2014,
   author = {Manuel Laubach and Johannes Reuther and Ronny Thomale and Stephan Rachel},
   doi = {10.1103/PhysRevB.90.165136},
   issn = {1550235X},
   issue = {16},
   journal = {Phys. Rev. B},
   month = {10},
   publisher = {American Physical Society},
   title = {Rashba spin-orbit coupling in the {Kane-Mele-Hubbard} model},
   volume = {90},
   pages = {165136},
   year = {2014}
}

@article{Moore2010,
   author = {Joel E. Moore},
   doi = {10.1038/nature08916},
   issn = {00280836},
   issue = {7286},
   journal = {Nature},
   month = {3},
   pages = {194-198},
   pmid = {20220837},
   title = {The birth of topological insulators},
   volume = {464},
   year = {2010}
}

@article{doi.org/10.1002/adma.200600527,
author = {Dresselhaus, M.S. and Chen, G. and Tang, M.Y. and Yang, R.G. and Lee, H. and Wang, D.Z. and Ren, Z.F. and Fleurial, J.-P. and Gogna, P.},
title = {New Directions for Low-Dimensional Thermoelectric Materials},
journal = {Adv. Mater.},
volume = {19},
number = {8},
pages = {1043-1053},
doi = {https://doi.org/10.1002/adma.200600527},
url = {https://advanced.onlinelibrary.wiley.com/doi/abs/10.1002/adma.200600527},
year = {2007}
}

@article{Zhao2014,
   author = {Li Dong Zhao and Shih Han Lo and Yongsheng Zhang and Hui Sun and Gangjian Tan and Ctirad Uher and C. Wolverton and Vinayak P. Dravid and Mercouri G. Kanatzidis},
   doi = {10.1038/nature13184},
   issn = {14764687},
   issue = {7496},
   journal = {Nature},
   pages = {373-377},
   pmid = {24740068},
   publisher = {Nature Publishing Group},
   title = {Ultralow thermal conductivity and high thermoelectric figure of merit in {SnSe} crystals},
   volume = {508},
   year = {2014}
}

@article{Chadov2010,
   author = {Stanislav Chadov and Xiaoliang Qi and Jürgen Kübler and Gerhard H. Fecher and Claudia Felser and Shou Cheng Zhang},
   doi = {10.1038/nmat2770},
   issn = {14764660},
   issue = {7},
   journal = {Nature Mater.},
   pages = {541-545},
   publisher = {Nature Publishing Group},
   title = {Tunable multifunctional topological insulators in ternary {Heusler} compounds},
   volume = {9},
   year = {2010}
}

@article{Casper2012,
   author = {F. Casper and T. Graf and S. Chadov and B. Balke and C. Felser},
   doi = {10.1088/0268-1242/27/6/063001},
   issn = {02681242},
   issue = {6},
   journal = {Semicond. Sci. Technol.},
   title = {Half-{Heusler} compounds: Novel materials for energy and spintronic applications},
   volume = {27},
   pages = {063001},
   year = {2012}
}

@article{Liu2025,
   author = {Wei Liu and Tao Hong and Xinxiu Cheng and Liqing Xu and Guilong Yan and Wenke He and Yu Xiao},
   doi = {10.1016/j.jmat.2024.100938},
   issn = {23528486},
   issue = {4},
   journal = {J. Materiomics},
   month = {7},
   publisher = {Chinese Ceramic Society},
   title = {Alloy scattering to optimize carrier and phonon transport properties in {PbBi2S4} thermoelectric},
   volume = {11},
   pages = {100938},
   year = {2025}
}

@article{Verma2023,
   author = {Ajay Kumar Verma and Shamma Jain and Kishor Kumar Johari and Christophe Candolfi and Bertrand Lenoir and Sumeet Walia and S. R. Dhakate and Bhasker Gahtori},
   doi = {10.1039/d3ma00923h},
   issn = {26335409},
   issue = {24},
   journal = {Mater. Adv.},
   month = {11},
   pages = {6655-6664},
   publisher = {Royal Society of Chemistry},
   title = {Approaching the minimum lattice thermal conductivity in {TiCoSb} half-{Heusler} alloys by intensified point-defect phonon scattering},
   volume = {4},
   year = {2023}
}

@article{Zhu2023,
   author = {Hangtian Zhu and Wenjie Li and Amin Nozariasbmarz and Na Liu and Yu Zhang and Shashank Priya and Bed Poudel},
   doi = {10.1038/s41467-023-38446-0},
   issn = {20411723},
   issue = {1},
   journal = {Nat. Commun.},
   month = {12},
   pmid = {37280195},
   publisher = {Nature Research},
   title = {Half-{Heusler} alloys as emerging high power density thermoelectric cooling materials},
   volume = {14},
   pages = {3300},
   year = {2023}
}

@article{Mitra2022,
   author = {Mousumi Mitra and Allen Benton and Md Sabbir Akhanda and Jie Qi and Mona Zebarjadi and David J. Singh and S. Joseph Poon},
   doi = {10.1016/j.mtphys.2022.100900},
   issn = {25425293},
   journal = {Mater. Today Phys.},
   month = {11},
   publisher = {Elsevier Ltd},
   title = {Conventional Half-{Heusler} alloys advance state-of-the-art thermoelectric properties},
   volume = {28},
   pages = {100900},
   year = {2022}
}

@article{Mozharivskii1998,
   author = {A. Mozharivskii and Y. B. Kuz’ma and O. M. Sichevich},
   issue = {7},
   journal = {Inorg. Mater.},
   pages = {707-710},
   title = {{Y-Ni-Bi} and {Ho-Ni-Bi} systems},
   volume = {34},
   year = {1998}
}

@article{Levin2016,
   author = {E. M. Levin},
   doi = {10.1103/PhysRevB.93.245202},
   issn = {24699969},
   issue = {24},
   journal = {Phys. Rev. B},
   month = {6},
   publisher = {American Physical Society},
   title = {Charge carrier effective mass and concentration derived from combination of Seebeck coefficient and Te 125 NMR measurements in complex tellurides},
   volume = {93},
   pages = {245202},
   year = {2016}
}

@article{Pagliuso,
   author = {P G Pagliuso and C Rettori and M E Torelli and G B Martins and Z Fisk and J L Sarrao and M F Hundley and S B Oseroff},
   title = {Crystal-field study in rare-earth-doped semiconducting {YBiPt}},
   doi = {10.1103/PhysRevB.60.4176},
   journal = {Phys. Rev. B},
   volume = {60},
   pages = {4176},
   year = {1999}
}

@article{Min2024,
   author = {Ruonan Min and Yanxia Wang and Xue Jiang and Rongchun Chen and Mingyang Li and Huijun Kang and Xiong Yang and Zongning Chen and Enyu Guo and Tongmin Wang},
   doi = {10.1016/j.jmat.2023.05.009},
   issn = {23528486},
   issue = {1},
   journal = {J. Materiomics},
   month = {1},
   pages = {200-209},
   publisher = {Chinese Ceramic Society},
   title = {{ZrNiSn}-based compounds with high thermoelectric performance and ultralow lattice thermal conductivity via introduction of multiscale scattering centers},
   volume = {10},
   pages = {200-209},
   year = {2024}
}

@article{Barczak2018,
   author = {Sonia A. Barczak and Jim Buckman and Ronald I. Smith and Annabelle R. Baker and Eric Don and Ian Forbes and Jan Willem G. Bos},
   doi = {10.3390/ma11040536},
   issn = {19961944},
   issue = {4},
   journal = {Materials},
   month = {3},
   publisher = {MDPI AG},
   title = {Impact of interstitial {Ni} on the thermoelectric properties of the half-{Heusler} {TiNiSn}},
   volume = {11},
   pages = {536},
   year = {2018}
}

@article{Sekimoto2006,
   author = {Takeyuki Sekimoto and Ken Kurosaki and Hiroaki Muta and Shinsuke Yamanaka},
   doi = {10.1016/j.jallcom.2005.06.036},
   issn = {09258388},
   issue = {1-2},
   journal = {J. Alloys Compd.},
   month = {1},
   pages = {326-329},
   title = {Thermoelectric properties of {Sn}-doped {TiCoSb} half-{Heusler} compounds},
   volume = {407},
   pages = {326-329},
   year = {2006}
}

@article{Ouardi2011,
   author = {Siham Ouardi and Gerhard H. Fecher and Claudia Felser and Jaroslav Hamrle and Kamil Postava and Jaromr Piŝtora},
   doi = {10.1063/1.3663569},
   issn = {00036951},
   issue = {21},
   journal = {Appl. Phys. Lett.},
   month = {11},
   title = {Transport and optical properties of the gapless {{Heusler}} compound {PtYSb}},
   volume = {99},
   pages = {211904},
   year = {2011}
}

@article{Chauhan2022,
   author = {Nagendra S. Chauhan and Yuzuru Miyazaki},
   doi = {10.1016/j.mtla.2022.101430},
   issn = {25891529},
   journal = {Materialia},
   month = {5},
   publisher = {Elsevier B.V.},
   title = {Low lattice thermal conductivity and microstructural evolution in {VFeSb} half-{Heusler} alloys},
   volume = {22},
   pages = {101430},
   year = {2022}
}

@article{Zhao2018,
   author = {Degang Zhao and Min Zuo and Lin Bo and Yongpeng Wang},
   doi = {10.3390/ma11050728},
   issn = {19961944},
   issue = {5},
   journal = {Materials},
   month = {5},
   publisher = {MDPI AG},
   title = {Synthesis and thermoelectric properties of Pd-Doped {ZrCoBi} Half-{Heusler} compounds},
   volume = {11},
   pages = {728},
   year = {2018}
}

@article{Chauhan2019,
   author = {Nagendra S. Chauhan and Sivaiah Bathula and Avinash Vishwakarma and Ruchi Bhardwaj and Kishor Kumar Johari and Bhasker Gahtori and Ajay Dhar},
   doi = {10.1016/j.jmat.2018.11.003},
   issn = {23528486},
   issue = {1},
   journal = {J. Materiomics},
   month = {3},
   pages = {94-102},
   publisher = {Chinese Ceramic Society},
   title = {Enhanced thermoelectric performance in p-type {ZrCoSb} based half-{Heusler} alloys employing nanostructuring and compositional modulation},
   volume = {5},
   year = {2019}
}

@article{Kahiu2023,
   author = {Joseph Ngugi Kahiu and Samuel Kimani Kihoi and Hyunji Kim and U. Sandhya Shenoy and D. Krishna Bhat and Ho Seong Lee},
   doi = {10.1021/acsaem.3c00253},
   issn = {25740962},
   issue = {8},
   journal = {ACS Appl. Energy Mater.},
   month = {4},
   pages = {4305-4316},
   publisher = {American Chemical Society},
   title = {Asymmetric Thermoelectric Performance Tuning in Low-Cost {ZrFe\textsubscript{x}Ni\textsubscript{1-x}Sb {Double Half-{Heusler} Materials}}},
   volume = {6},
   year = {2023}
}

@article{PhysRevB.50.17953,
  title = {Projector augmented-wave method},
  author = {Bl\"ochl, P. E.},
  journal = {Phys. Rev. B},
  volume = {50},
  issue = {24},
  pages = {17953--17979},
  numpages = {0},
  year = {1994},
  month = {Dec},
  publisher = {American Physical Society},
  doi = {10.1103/PhysRevB.50.17953},
  url = {https://link.aps.org/doi/10.1103/PhysRevB.50.17953}
}

@article{PhysRevB.59.1758,
  title = {From ultrasoft pseudopotentials to the projector augmented-wave method},
  author = {Kresse, G. and Joubert, D.},
  journal = {Phys. Rev. B},
  volume = {59},
  issue = {3},
  pages = {1758--1775},
  numpages = {0},
  year = {1999},
  month = {Jan},
  publisher = {American Physical Society},
  doi = {10.1103/PhysRevB.59.1758},
  url = {https://link.aps.org/doi/10.1103/PhysRevB.59.1758}
}

@article{PhysRevB.47.558,
  title = {Ab initio molecular dynamics for liquid metals},
  author = {Kresse, G. and Hafner, J.},
  journal = {Phys. Rev. B},
  volume = {47},
  issue = {1},
  pages = {558(R)--561(R)},
  numpages = {0},
  year = {1993},
  month = {Jan},
  publisher = {American Physical Society},
  doi = {10.1103/PhysRevB.47.558},
  url = {https://link.aps.org/doi/10.1103/PhysRevB.47.558}
}

@article{PhysRevB.54.11169,
  title = {Efficient iterative schemes for ab initio total-energy calculations using a plane-wave basis set},
  author = {Kresse, G. and Furthm\"uller, J.},
  journal = {Phys. Rev. B},
  volume = {54},
  issue = {16},
  pages = {11169--11186},
  numpages = {0},
  year = {1996},
  month = {Oct},
  publisher = {American Physical Society},
  doi = {10.1103/PhysRevB.54.11169},
  url = {https://link.aps.org/doi/10.1103/PhysRevB.54.11169}
}

@article{KRESSE199615,
title = {Efficiency of ab-initio total energy calculations for metals and semiconductors using a plane-wave basis set},
journal = {Comput. Mater. Sci.},
volume = {6},
number = {1},
pages = {15-50},
year = {1996},
issn = {0927-0256},
doi = {https://doi.org/10.1016/0927-0256(96)00008-0},
url = {https://www.sciencedirect.com/science/article/pii/0927025696000080},
author = {G. Kresse and J. Furthmüller}
}

@article{PhysRevB.49.14251,
  title = {Ab initio molecular-dynamics simulation of the liquid-metal--amorphous-semiconductor transition in germanium},
  author = {Kresse, G. and Hafner, J.},
  journal = {Phys. Rev. B},
  volume = {49},
  issue = {20},
  pages = {14251--14269},
  numpages = {0},
  year = {1994},
  month = {May},
  publisher = {American Physical Society},
  doi = {10.1103/PhysRevB.49.14251},
  url = {https://link.aps.org/doi/10.1103/PhysRevB.49.14251}
}

@article{PhysRevLett.77.3865,
  title = {Generalized Gradient Approximation Made Simple},
  author = {Perdew, John P. and Burke, Kieron and Ernzerhof, Matthias},
  journal = {Phys. Rev. Lett.},
  volume = {77},
  issue = {18},
  pages = {3865--3868},
  numpages = {0},
  year = {1996},
  month = {Oct},
  publisher = {American Physical Society},
  doi = {10.1103/PhysRevLett.77.3865},
  url = {https://link.aps.org/doi/10.1103/PhysRevLett.77.3865}
}

@article{PhysRevB.59.15660,
  title = {Electronic structure of rare-earth nickel pnictides: Narrow-gap thermoelectric materials},
  author = {Larson, P. and Mahanti, S. D. and Sportouch, Sandrine and Kanatzidis, M. G.},
  journal = {Phys. Rev. B},
  volume = {59},
  issue = {24},
  pages = {15660--15668},
  numpages = {0},
  year = {1999},
  month = {Jun},
  publisher = {American Physical Society},
  doi = {10.1103/PhysRevB.59.15660},
  url = {https://link.aps.org/doi/10.1103/PhysRevB.59.15660}
}

@article{https://doi.org/10.1002/adfm.200701369,
author = {Yang, Jiong and Li, Huanming and Wu, Ting and Zhang, Wenqing and Chen, Lidong and Yang, Jihui},
title = {Evaluation of Half-{Heusler} Compounds as Thermoelectric Materials Based on the Calculated Electrical Transport Properties},
journal = {Adv. Funct. Mater.},
volume = {18},
number = {19},
pages = {2880-2888},
doi = {https://doi.org/10.1002/adfm.200701369},
year = {2008}
}

@article{PhysRevLett.109.037602,
  title = {Half-{Heusler} Semiconductors as Piezoelectrics},
  author = {Roy, Anindya and Bennett, Joseph W. and Rabe, Karin M. and Vanderbilt, David},
  journal = {Phys. Rev. Lett.},
  volume = {109},
  issue = {3},
  pages = {037602},
  numpages = {5},
  year = {2012},
  month = {Jul},
  publisher = {American Physical Society},
  doi = {10.1103/PhysRevLett.109.037602},
  url = {https://link.aps.org/doi/10.1103/PhysRevLett.109.037602}
}

@article{10.1063/1.4799144,
    author = {Shekhar, Chandra and Nicklas, M. and Nayak, Ajaya K. and Ouardi, Siham and Schnelle, W. and Fecher, Gerhard H. and Felser, Claudia and Kobayashi, Keisuke},
    title = {Electronic structure and nonsaturating magnetoresistance of superconducting {Heusler} topological insulators},
    journal = {J. Appl. Phys.},
    volume = {113},
    number = {17},
    pages = {17E142},
    year = {2013},
    month = {04},
    issn = {0021-8979},
    doi = {10.1063/1.4799144}
}

@article{Heinrich_Hohl_1999,
doi = {10.1088/0953-8984/11/7/004},
url = {https://doi.org/10.1088/0953-8984/11/7/004},
year = {1999},
month = {feb},
publisher = {},
volume = {11},
number = {7},
pages = {1697},
author = {Heinrich Hohl and Art P Ramirez and Claudia Goldmann and Gabriele Ernst and Bernd Wölfing and Ernst Bucher},
title = {Efficient dopants for {ZrNiSn-based} thermoelectric materials},
journal = {J. Phys.: Condens. Matter}
}

\end{document}